\documentstyle[12pt,epsfig]{article}
\def\vpy#1#2#3{{\bf #1}, #2 (#3)}
\begin{document}
\begin{center}{\Large
A total neutrino conversion in the Earth\\
\vspace{3mm}
without a resonance.}\\
\ \\
M. V. Chizhov\footnote{Permanent address: Centre for Space Research
and Technologies,
Faculty of Physics,\\ University of Sofia, 1164 Sofia, Bulgaria}\\
\ \\
{\it The Abdus Salam International Centre for Theoretical Physics,\\
Strada Costiera 11, 34014 Trieste, Italy}
\end{center}
\ \\
\begin{abstract}
The neutrino oscillation enhancement in the Earth-type medium
{\it mantle -- core -- mantle} is discussed. It is noted 
that the total conversion is possible both for a resonant matter density
and a nonresonant one. A useful parameterization, for the representation
of the transition probability for neutrinos and antineutrinos in
a single plot, is proposed.
\end{abstract}
\ \\

The matter effect of the Earth on neutrino oscillations, in the 
interesting region of oscillation parameters for the solar and atmospheric
neutrinos,
is widely discussed now \cite{LS,P,ADLS,CMP,S,PRL,hep}.
Due to the specific multilayer structure of the Earth,
a new effect~\cite{PRL,hep} of an enhancement of the oscillations of
massive neutrinos is possible.
In contrast to MSW {\it resonance} effect \cite{MSW} this 
new effect occurs due to a
maximal {\it constructive interference} among transition amplitudes,
which give contribution to the total amplitude in the multilayer case.

A good approximation for the Earth interior is a two-layer model with
two basic structures: the mantle and the core.
These structures have slowly increasing
densities from the surface of the Earth to its center with a sharp leap on
their border. Therefore, we can consider the mantle and the core
densities on the neutrino trajectories as different constants. 
This assumption leads to simple analytical formulae
for the neutrino transition probability (see, for example, \cite{P}).

In the case of oscillations between
two ultra-relativistic neutrino species in vacuum, there are two
parameters: the vacuum mixing angle $\vartheta_0$ and the ratio
between the neutrino squared mass difference and the neutrino energy 
$\Delta m^2/E$. We assume for definiteness that $\Delta m^2 > 0$ and 
\begin{equation}
0 < \vartheta_0 \le \pi/4,
\label{angle}
\end{equation}
i.e.\ $\cos(2\vartheta_0) > 0$.
If the vacuum mixing angle $\vartheta_0$ is small, 
the neutrino transition probability is suppressed.
Therefore, in this case the experimental search of oscillations
is extremely difficult.

On the other hand, a medium can effect oscillations and, in particular,
enhance them. For oscillations in a medium, a difference
$V_{\alpha\beta}$ ($\alpha \ne \beta = e,\mu,\tau,s$) between the
effective potentials of different neutrino species
$\nu_\alpha$ and $\nu_\beta$ can arise.
For neutrinos it can be either positive
\begin{equation}
V_{e\mu}=\sqrt{2}G_F N_e > 0,
\label{emu}
\end{equation}
or negative
\begin{equation}
V_{\mu s}=-\sqrt{2}G_F N_n/2 < 0,
\label{mus}
\end{equation}
where $N_e$ and $N_n$ are the electron and neutron number densities of
the medium. For antineutrinos $V_{\alpha\beta}$ is replaced by
\begin{equation}
V_{\bar{\alpha}\bar{\beta}}=-V_{\alpha\beta}.
\label{anti}
\end{equation}  

The matter mixing angle $\vartheta$ is given by the well-known
expression
\begin{equation}
\cos(2\vartheta)={1 \over \Delta E}\left(
{\Delta m^2 \over 2E}\cos(2\vartheta_0)-V_{\alpha\beta}\right),
\end{equation} 
where
\begin{equation}
\Delta E={\Delta m^2 \over 2E}
\sqrt{\left(\cos(2\vartheta_0)-{2EV_{\alpha\beta}
\over \Delta m^2}\right)^2
+\sin^2(2\vartheta_0)}
\end{equation}
being the difference between the energies of the two neutrino
energy-eigenstates in the medium. 

In order to present a comparison of
the probabilities of neutrinos and antineutrinos
in a single figure, we extend formally the range of the vacuum mixing
angles (\ref{angle}) to the region 
\begin{equation}
0 < \vartheta_0 < \pi/2
\label{ext}
\end{equation}
in such a way that $\cos(2\vartheta_0)<0$ corresponds to the antineutrino
case, keeping the same $V_{\alpha\beta}$ for neutrinos and antineutrinos. 
This is possible, because in the two-species case we can 
obtain antineutrino evolution equation from the neutrino one
\begin{equation}
i{{\rm d}\over {\rm d}t}
\left(\begin{array}{c} \nu_\alpha\\ \nu_\beta
\end{array}\right) =
{\Delta E \over 2}
\left(\begin{array}{cc}
-\cos(2\vartheta) & \sin(2\vartheta)\\
\sin(2\vartheta) & \cos(2\vartheta)
\end{array}\right)
\left(\begin{array}{c} \nu_\alpha\\ \nu_\beta
\end{array}\right)
\end{equation}
by the formal substitutions:
$\cos(2\vartheta_0) \to -\cos(2\vartheta_0)$ and 
$\nu_\alpha \to \bar{\nu}_\beta$, $\nu_\beta \to \bar{\nu}_\alpha$.
As far as $P_{\alpha\beta(\bar{\alpha}\bar{\beta})}=
P_{\beta\alpha(\bar{\beta}\bar{\alpha})}$, 
we can plot the {\it continuous} total transition
probability $P_{\alpha\beta}$ for neutrinos and antineutrinos
in a single figure,
using extended region (\ref{ext}) for the vacuum mixing angles.

When the MSW resonance condition
\begin{equation}
{\Delta m^2 \over 2E}\cos(2\vartheta_0)=V_{\alpha\beta}
\label{res}
\end{equation}
is fulfilled, the matter mixing angle can be maximal, $\vartheta=\pi/4$,
even in the case of a small vacuum mixing angle $\vartheta_0$.
In this case the neutrino transition probability can reach
its maximal value $P_{\alpha\beta}=1$. 
It can be realized either for neutrinos 
or for antineutrinos. In a constant density
{\it homogeneous} medium the maxima of
neutrino transition probability lie on the curve~(\ref{res}) in
$(\cos(2\vartheta_0),\Delta m^2/E)-$plane. 
The positions of the maxima depend on the distance $X$, travelled
by the neutrinos or antineutrinos, and
are defined by the phase condition
\begin{equation}
\phi=\Delta E X=(2k+1)\pi,~~~~k=0,1,2,\dots~.
\label{phase} 
\end{equation}

When the (anti)neutrinos arrive to the detector 
at nadir angle greater than
$33^\circ$, they pass only through the Earth mantle, which 
is assumed to have a constant density $\rho_m\cong 4.5$ g/cm$^3$. 
Therefore, we can consider it as a simple case of a neutrino propagation
in a constant density homogeneous medium. The contours of the 
analytically calculated transition
probability, at the nadir angle $h=70^\circ$ for different oscillation
parameters in the cases of 
$\stackrel{(-)}{\nu}_\mu \leftrightarrow 
\stackrel{(-)}{\nu}_\tau$ and 
$\stackrel{(-)}{\nu}_\mu \leftrightarrow \stackrel{(-)}{\nu}_s$ 
oscillations, are shown in Fig.~1.
In the case of $\stackrel{(-)}{\nu}_\mu \leftrightarrow 
\stackrel{(-)}{\nu}_\tau$
oscillations, $V_{\mu\tau}=0$. This corresponds to a vacuum case, when
a total conversion $P_{\mu\tau}=1$ takes place only at a maximal 
vacuum mixing
angle $\vartheta_0=\pi/4$, i.e.\ $\cos(2\vartheta_0)=0$ (Fig.~1a).
The transition probability has a symmetrical form and there is no
difference between neutrino and antineutrino cases.
The later case of $\stackrel{(-)}{\nu}_\mu \leftrightarrow 
\stackrel{(-)}{\nu}_s$ oscillations allows
a total resonance conversion for antineutrino
oscillations $\bar{\nu}_\mu \leftrightarrow \bar{\nu}_s$, and 
a supression of the transition probability for 
$\nu_\mu \leftrightarrow \nu_s$ case (Fig.~1b).
This feature enables us to distinguish between
these cases for the atmospheric neutrinos.
In the following we consider just the latter case which is effected
by a matter distribution.

In the case of smaller nadir angles $h\le 33^\circ$, (anti)neutrinos
pass also through the Earth core, which density we assume to be 
constant $\rho_c\cong 11.5$ g/cm$^3$. This leads to simple analytical
expressions for the neutino transition probability, which has been
analyzed in \cite{hep}. Due to the strong
interference between the amplitudes in the mantle and the core,
a total conversion 
{\it even} for neutrinos $P_{\mu s}=1$ can occur. At the nadir
angles just below 33$^\circ$ the absolute maxima 
move away from the curve (\ref{res})
and their interpretation in the terms of the MSW resonances becames
meaningless. As it was shown in \cite{hep}, in the three-layer case of
the Earth profile, the total (anti)neutrino conversion is possible
in the infinite two-dimensional region of the oscillation parameters
\begin{equation}
region~{\cal A}:~
\left\{
\begin{array}{l}
\cos(2\vartheta_c) \le 0 \\
\cos(2\vartheta_c-4\vartheta_m) \ge 0,
\end{array} \right.
\label{A}
\end{equation}
where $\vartheta_m$ and $\vartheta_c$ are the mixing angles 
in the mantle and the core, correspondingly.
For fixed $\Delta m^2$ and the vacuum mixing angle $\vartheta_0$, the
conditions
(\ref{A}) give the allowed values of neutrino energy $E$, at which a total
conversion in the Earth is possible.
In contrast to the MSW resonance condition (\ref{res}), where only single
value of $E$ is possible, there exists a continuum of different solutions
for $E$. 
Moreover, the region ${\cal A}$ is wider than the analogous region in 
the two-layer case considered in \cite{PRL}. For $\rho_c>2\rho_m$,
which has place for the Earth, the region ${\cal A}$ overlaps
the parameter space $\cos(2\vartheta_0)>0$,
where the MSW resonance condition cannot be satisfied, due to 
the different signs on the left and on the right hand side 
of eq.~(\ref{res}). 
However, a total neutrino conversion is possible. It is somewhat opposite
to the common opinion that in this case the matter suppresses oscillations
and the total neutrino conversion cannot occur.

The positions of the absolute maxima in the region ${\cal A}$
are defined by the two conditions on the phases in the mantle $\phi_m$
and in the core $\phi_c$
\begin{equation}   
\left\{
\begin{array}{l}
\tan{\displaystyle{\phi_m\over 2}}
=\pm\sqrt{{\displaystyle -\cos(2\vartheta_c)\over
\displaystyle\cos(2\vartheta_c- 4\vartheta_m)}},
\\
\tan{\displaystyle{\phi_c\over 2}}
=\pm{\displaystyle \cos(2\vartheta_m)\over \sqrt{
\displaystyle-\cos(2\vartheta_c)\cos(2\vartheta_c- 4\vartheta_m)}}.
\end{array} \right.
\label{Earth}
\end{equation}
In Fig.~2, for example,
we show the contours of the transition probability at
nadir angle $h=32.4^\circ$. The rightmost maximum corresponds to
a total neutrino conversion into nonresonance region.

The total conversion can
take place for $\nu_\mu \leftrightarrow \nu_s$
oscillations near the maximal vacuum mixing angle 
$\sin^2(2\vartheta_0)>0.993$ and in the wide range
of the vacuum mixing angles for the resonant case of
$\bar{\nu}_\mu \leftrightarrow \bar{\nu}_s$ oscillations. It can lead 
to a specific dependence of 
the nadir angle distribution for the atmospheric
neutrinos. In \cite{LS} it was noted that at the maximal mixing angle
$\vartheta_0=\pi/4$, $\Delta m^2/E\cong 2\times 10^{-4}$ eV$^2$/GeV and
nadir angle near 30$^\circ$,
when the special conditions
\begin{equation}   
\phi_m=\phi_c=\pi
\label{NORL}  
\end{equation}
are approximately satisfied,
the enhancement of $\nu_\mu \leftrightarrow \nu_s$ oscillations
takes place. However,
the equalities (\ref{NORL}) are not the right conditions for the    
maximum of the transition probability, in contrast to the opinion   
of the authors of ref.~\cite{LS}.
According to our approach
these conditions correspond to the limiting case,
when the absolute maxima lie on the boundary
\begin{equation}
\cos(2\vartheta_c-4\vartheta_m)=0
\label{NORLcurve}
\end{equation}
of region ${\cal A}$.
This curve defines how far from the resonance curve (\ref{res}) the total
neutrino conversion for the Earth-type profile can occur. 
The enhancement found in \cite{LS} 
is due to the lowest absolute maximum, which is a solution of
eqs.~(\ref{Earth}), near to the maximal mixing angle (see Fig.~2).

For nonresonant matter oscillations the region, where
the total neutrino conversion occurs, becomes maximal in the case of
the {\it vacuum -- matter -- vacuum} profile (Fig.~3). 
The minimal possible
vacuum mixing angle, at which the total neutrino conversion takes place,
is equal to $\pi/8$, i.e.\ $\sin^2(2\vartheta_0)\ge 1/2$. It has
a clear physical meaning: the role of the inner layer is to prevent 
the rapid decrease of the transition probability,
after it reaches its maximal value in the first layer (Fig.~4).
Therefore, two outer layers are enough for the realization of a
total neutrino conversion.

The limiting case of small mixing angles and `optimal' conditions
on the phases (\ref{phase},\ref{NORL})
is analogous to the parametric enhancement
of oscillations considered in \cite{par},
where `drift' of the transition probability to its maximal
value takes place.
In these cases, when oscillation amplitudes are small, many periods
in a medium with periodic number density are required to reach the
absolute maximum of the transition probability $P=1$.
As far as in the nonresonance region
the area, where the total neutrino conversion occurs,
extends more and more with each period
to small mixing angles.
However, the authors of refs. \cite{par,A} have missed all solutions 
for absolute maxima of the transition probability
inside this area.
For the Earth profile, for instance, they correspond to 
conditions on phases, which depend on the matter angles
$\vartheta_m$ and $\vartheta_c$ (see \ref{Earth}).
It was shown, that without the assumption about
the small vacuum mixing
angles or small matter effect, the transition probability can
reach its absolute maximum $P=1$ for the three-layer Earth-type
medium and {\it even} for the two-layer case \cite{PRL,hep}.

So, the total conversion in a multilayer medium can occur both
in the resonance and the nonresonance regions of oscillation parameters.
This simple fact must be kept in mind, when the neutrino propagation in 
a multilayer medium with different densities like the Earth is analyzed.

I express my gratitude to SISSA and ICTP for the warm
hospitality and financial support. I am also obliged to S. T. Petcov
for introducing me into this theme, fruitful collaboration and help. 
I thank Q. Y. Liu for useful discussions.

\pagebreak[1]

\newpage
\pagestyle{empty}   

{\bf Figure captions}\\

{\bf Figure 1a.} The contours for the different values of the transition
probability $P_{\mu\tau}$ 0.2, 0.4, 0.6, 0.8 at nadir angle $h=70^\circ$
are shown. The dark spots inside them correspond to the absolute maxima
$P_{\mu\tau}=1$, which are realized at the maximal vacuum mixing angle
$\vartheta_0=\pi/4$.\\

{\bf Figure 1b.} The contours for the different values of the transition
probability $P_{\mu s}$ 0.2, 0.4, 0.6, 0.8 at nadir angle $h=70^\circ$
are shown. The dark spots inside them correspond to the absolute maxima
$P_{\mu s}=1$. The resonance curve for the mantle,
where the total conversion can occur, is also drawn.\\

{\bf Figure 2.} The contours for the different values of the transition
probability $P_{\mu s}$ 0.2, 0.4, 0.6, 0.8 at nadir angle $h=32.4^\circ$
are shown. The dark spots inside them correspond to the absolute maxima
$P_{\mu s}=1$. The region ${\cal A}$, where the total conversion can
occur, is also drawn. For comparison the resonance curve for the mantle
(dot curve) is presented.\\

{\bf Figure 3.} The region ${\cal A}$ for 
{\it vacuum -- matter -- vacuum} profile, where the total conversion 
$P_{\mu s}=1$ can occur, is plotted.\\

{\bf Figure 4.} The evolution of the transition probability $P_{\mu s}$
in the medium {\it vacuum -- matter -- vacuum} at 
$\sin^2(2\vartheta_0)=1/2$ and $\Delta m^2/E \to 0$ is presented.

\newpage
\pagestyle{empty}
\begin{figure}   
\epsfig{file=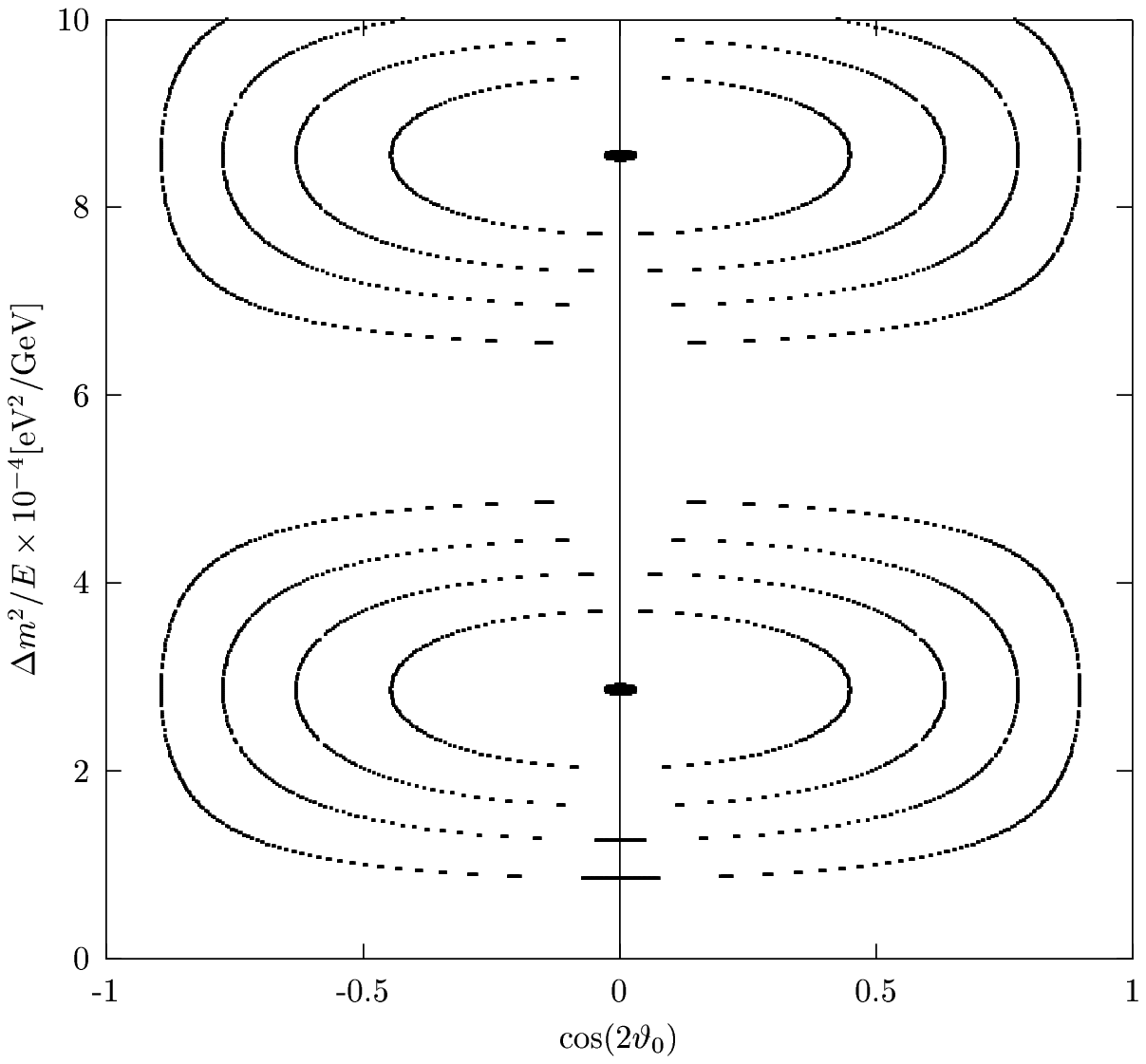,height=13cm,width=13cm}
\end{figure}

\begin{center}
Figure 1a.
\end{center}

\newpage
\pagestyle{empty}
\begin{figure}   
\epsfig{file=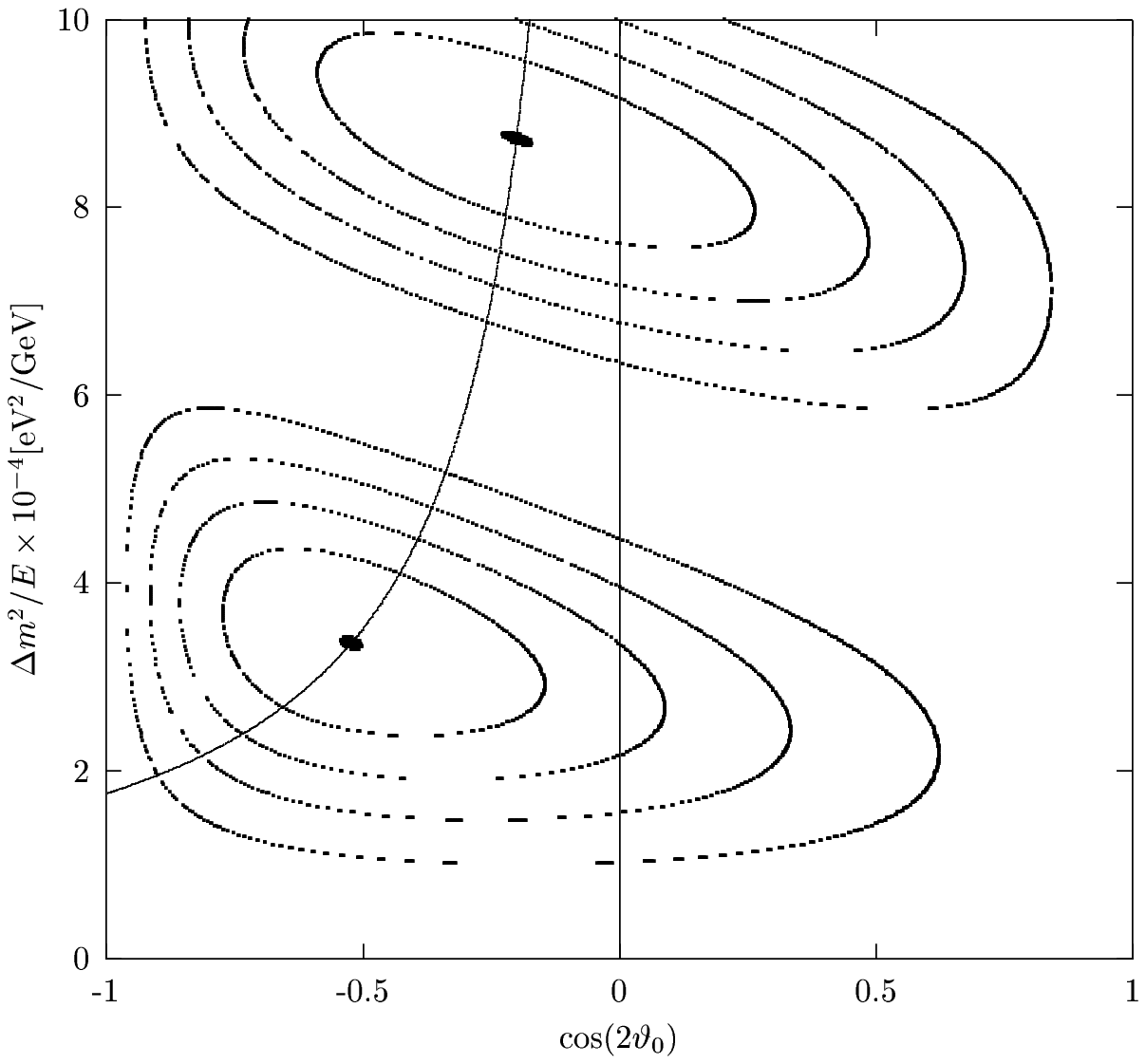,height=13cm,width=13cm}
\end{figure}

\begin{center}
Figure 1b.  
\end{center}

\newpage
\pagestyle{empty}
\begin{figure}   
\epsfig{file=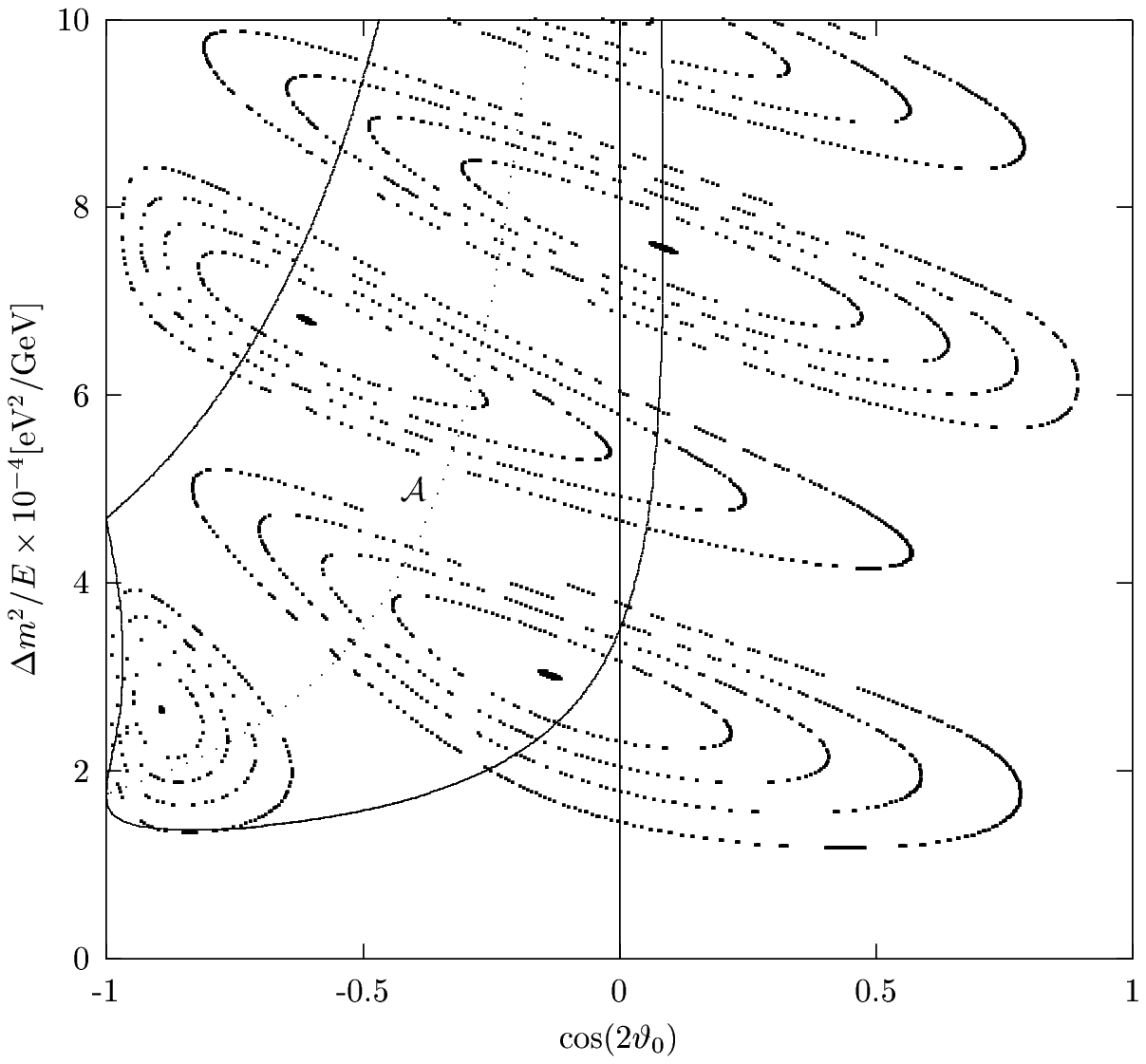,height=13cm,width=13cm}
\end{figure}

\begin{center}
Figure 2.
\end{center}

\newpage
\pagestyle{empty}
\begin{figure}   
\epsfig{file=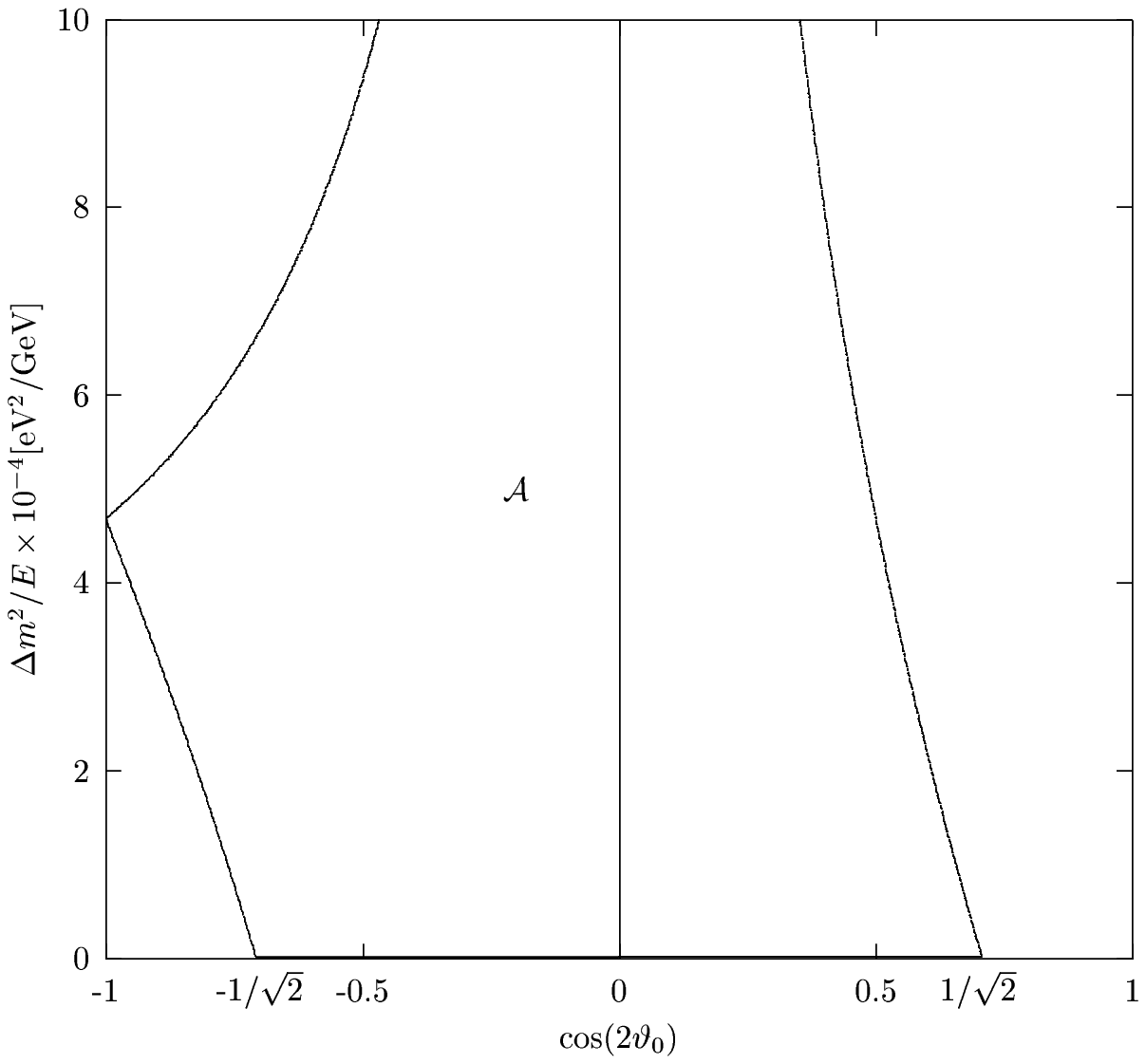,height=13cm,width=13cm}
\end{figure}

\begin{center}
Figure 3.
\end{center}

\newpage
\pagestyle{empty}
\begin{figure}
\epsfig{file=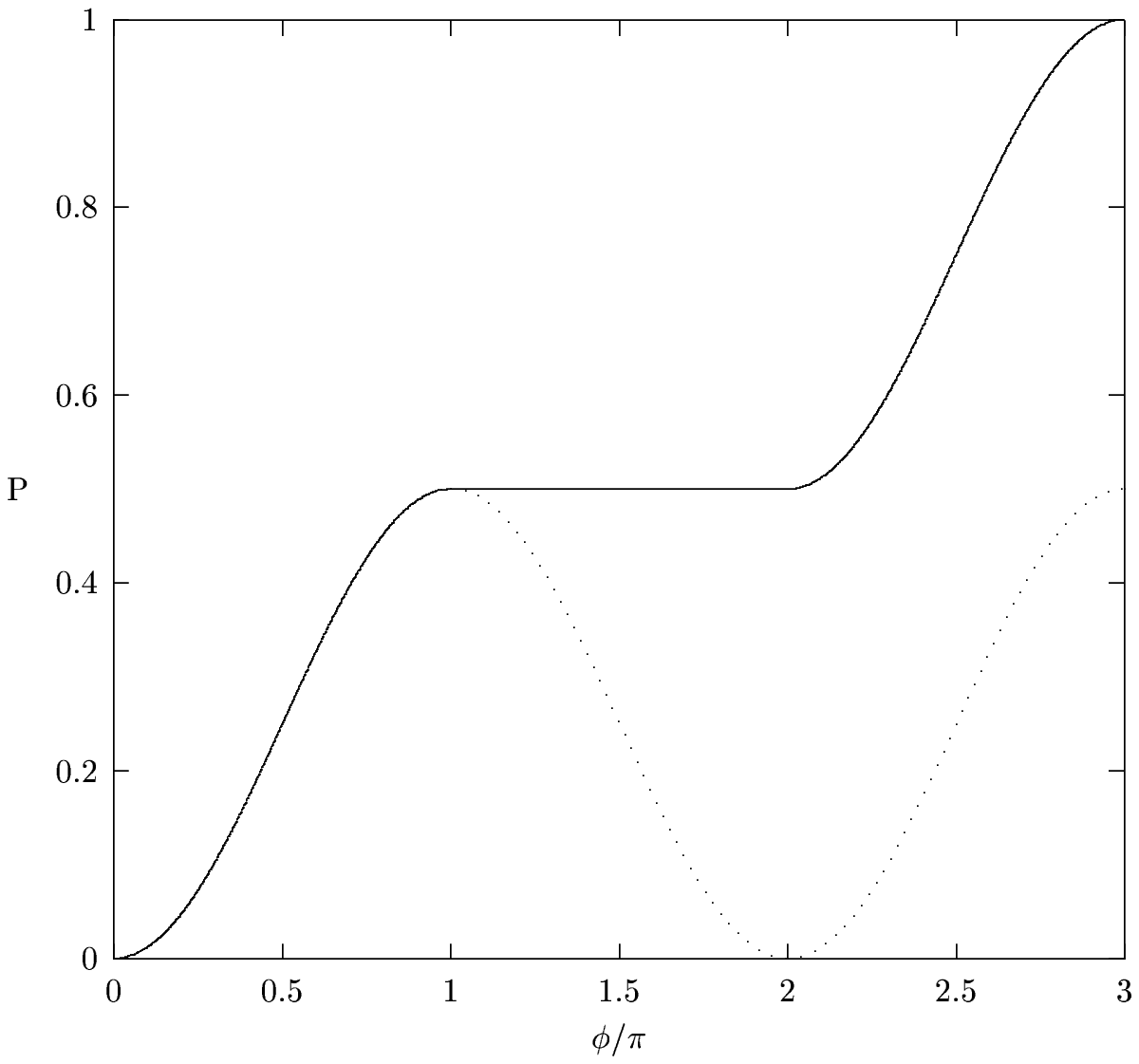,height=13cm,width=13cm}
\end{figure}

\begin{center}
Figure 4.
\end{center}

\end{document}